\documentclass[
aps,prl,reprint,
superscriptaddress,
amsmath,amssymb
]{revtex4-2}
\usepackage{graphicx}
\usepackage{amsmath,amssymb,bm}
\usepackage{mathrsfs}
\usepackage[T1]{fontenc}
\usepackage[utf8]{inputenc}
\usepackage{lmodern}
\usepackage[margin=27mm]{geometry}
\usepackage{amsmath,amssymb,bm}
\usepackage{microtype}
\usepackage[hidelinks]{hyperref}
\allowdisplaybreaks
\numberwithin{equation}{section}
\usepackage{physics}
\usepackage{relsize}
\usepackage{tikz}
\usepackage{tikz-cd}
\usepackage{hyperref}
\usepackage{booktabs}
\usetikzlibrary{arrows.meta, positioning}
\usepackage{ifthen}
\usepackage{pgf}
\usepackage{pgfmath}
\usepackage[utf8]{inputenc}
\usepackage{amsmath,amssymb,amsfonts}
\usepackage{geometry}
\usepackage{hyperref}
\usepackage{enumitem}
\usepackage{tikz}
\usepackage{animate}
\usepackage{ifthen}
\usepackage{graphicx}
\usepackage{booktabs}
\usepackage{booktabs}
\usepackage{bm}
\begin{document}

\title{
Emergent Spinorial Structure in Mesoscopic Irreversible Thermodynamics
}

\author{Giorgio Sonnino}

\affiliation{
Universit\'e libre de Bruxelles (ULB),
1050 Brussels, Belgium
}

\affiliation{
International Solvay Institutes for
Physics and Chemistry,
1050 Brussels, Belgium
}

%\date{\today}

% ============================================================
% ABSTRACT
% ============================================================

\begin{abstract}

Can the transport coefficients of a mesoscopic system determine both its coherent evolution and the geometric distance between its thermodynamic states? This work establishes that connection within a canonical operator framework for thermodynamic variables. We assume that thermodynamic coordinates and their conjugate forces satisfy commutation relations governed by a reduced constant ${\slash\!\!\!k}_B$. Requiring coordinate evolution to reproduce self-adjoint flux observables then determines a compatible evolution generator up to an arbitrary multiplication term, which we set to zero in the minimal realization. The resulting dynamic equation governs the coherent evolution of mesoscopic thermodynamic amplitudes, with transport coefficients determining its coefficients and ${\slash\!\!\!k}_B$ setting the evolution scale. An exact first-order factorization enables us to reconstruct a geometry directly from the same generator through Connes’ spectral-distance prescription, including when the transport coefficients depend on the thermodynamic coordinates. The metric is twice the inverse of the symmetric transport matrix. Derivatives of the antisymmetric component with respect to thermodynamic coordinates can alter the dynamics while leaving distances unchanged, distinguishing the dynamical and geometric information carried by transport. Published thermoelectric data illustrate this geometric reconstruction. We also identify a possible experimental signature of ${\slash\!\!\!k}_B$: a separate quadratic entropy-production observable has adjacent entropy levels separated by this constant under the stated canonical and finite-mode matching assumptions. The framework thus connects canonical thermodynamic fluctuations, coherent dynamics, and transport geometry, and yields a conditional entropy-level prediction for experimental investigation. The evolution is unitary; deriving irreversible relaxation requires additional physical mechanisms, such as coarse-graining or environmental coupling.

\end{abstract}

\maketitle

% ============================================================
% INTRODUCTION
% ============================================================

\section{Introduction}\label{sec:introduction}

Transport coefficients tell us how readily heat and charge flow, and how one transport process influences another. Can they also determine the coherent evolution of a mesoscopic thermodynamic state and the distance between two such states? We address this question within a canonical operator description of thermodynamic variables. Under the assumptions specified below, we construct an evolution generator compatible with the flux observables and derive a metric from a first-order factorization of that same generator. We also examine which physical consequences depend on the reduced constant ${\slash\!\!\!k}_B$ entering the canonical relations. Classical irreversible thermodynamics connects forces, fluxes, and entropy production through transport coefficients \cite{Onsager1931a,Onsager1931b,deGroot1984}. At the mesoscopic scale, we adopt the canonical description introduced in Refs. \cite{Sonnino2024a,Sonnino2024b,Sonnino2025}. In the normalization used here, thermodynamic coordinates and their conjugate forces satisfy
\begin{equation}\label{I1}
[\hat{\alpha}\mu,\hat{X}^{\nu}]
=i\frac{{\slash\!\!\!k}_B}{2}\delta_\mu^\nu
\end{equation}
Here $\alpha_\mu$ denotes a thermodynamic coordinate and $X^\nu$ its conjugate force. The reduced constant $\slash!!!k_B$, distinct from the ordinary Boltzmann constant $k_B$, sets the scale of noncommutativity and the associated uncertainty bounds. We assume these commutators and investigate their consequences. Establishing their applicability to a particular physical system requires experimental assessment. Earlier studies proposed the canonical relations and examined entropy discretization, equilibrium matching, and the physical interpretation of the reduced constant \cite{Sonnino2024a,Sonnino2024b,Sonnino2025}. Those results provide the starting point for this paper. What remains to be constructed is a coherent evolution law consistent with specified mesoscopic fluxes, together with a geometric interpretation of the resulting operator. Related ideas also appear in equilibrium thermodynamics. Santos et al. develop a canonical quantization based on constrained systems, obtaining thermodynamic wave functions, uncertainty relations, and a Schrödinger-like equation with entropy as the evolution parameter \cite{Santos2026}. Our approach shares the use of noncommuting thermodynamic observables and builds on the earlier canonical mesoscopic formulation cited above. Here, however, we determine a generator for evolution in physical time by requiring compatibility with flux observables. We then extract a spectral distance from its exact first-order factorization. This connection between transport, coherent evolution, and geometry is the result developed below. Geometry itself is already familiar in transport theory. Thermodynamic Field Theory introduces a metric and an affine connection \cite{Sonnino2009,Sonnino2016,Sonnino2022,Sonnino2002,Sonnino2019,Sonnino2001}; in a different setting, a generalized friction tensor defines a thermodynamic metric for near-equilibrium driving \cite{SivakCrooks2012}. We examine how a metric follows from the canonical transport generator, allowing its coefficients to depend on thermodynamic coordinates. Retaining their derivatives proves important: some transport contributions affect the evolution without entering the distance. To construct the evolution law, we choose self-adjoint flux observables and require the time derivative of each coordinate operator to reproduce the corresponding flux. This compatibility condition determines a self-adjoint generator up to an additive real multiplication operator $U(\alpha)$. We set $U=0$ in the minimal realization. The resulting dynamic equation governs a mesoscopic amplitude $\psi(\alpha,t)$ in thermodynamic-coordinate space, with transport fixing the differential coefficients and ${\slash\!\!\!k}_B$ setting the evolution scale. The equation describes coherent, probability-preserving evolution. Its self-adjoint generator gives unitary dynamics, so irreversible relaxation does not follow from this equation alone. Environmental coupling or coarse-graining would supply additional physical mechanisms through which a dissipative description could emerge. We obtain the geometry through Connes’ spectral-distance prescription \cite{Connes1994,GraciaBondia2001}. A first-order operator bounds the local variation of functions on the state space. Among functions satisfying that bound, the largest difference between their values at two points defines the distance. We use this established mathematical construction, but derive the required first-order operator from the transport generator; no separate metric needs to be prescribed. For a positive Hermitian transport tensor $\tau=g+if$, with $g$ real and symmetric and $f$ real and antisymmetric, an exact factorization on an auxiliary scalar–vector space recovers the evolution generator on its scalar sector. The commutators of the associated first-order operator yield the line element
\begin{equation}\label{I2}
ds^2=2,d\boldsymbol{\alpha}^{T}g(\boldsymbol{\alpha})^{-1}d\boldsymbol{\alpha}
\end{equation}
Thus, twice the inverse of the symmetric transport matrix gives the metric. For constant coefficients, equal coordinate displacements along directions with larger transport eigenvalues have shorter distances. These are distances in thermodynamic-coordinate space, where coordinates may describe transferred charge and heat. They are not lengths in physical space. When the coefficients vary with the thermodynamic coordinates, the reconstructed metric can have curvature. Derivatives of the antisymmetric tensor can also modify the evolution generator, but they leave the metric unchanged. The same distances can therefore accompany different dynamics. Under the regularity and completeness conditions stated below, the spectral distance agrees with the shortest-path distance associated with the metric. Applying Connes’ prescription does not require the thermodynamic coordinates themselves to have discrete spectra. In the distance calculation, coordinate functions form a commutative multiplication algebra, even though coordinates and their conjugate forces obey noncommuting canonical relations. The reconstructed metric defines its own Levi-Civita connection. It does not determine the separate, generally non-Levi-Civita connection of Thermodynamic Field Theory. In the constant Onsager regime, a Clifford–Dirac representation gives the same distance; its spinorial components implement an algebra fixed by transport without introducing physical spin. The normalized metric does not explicitly contain ${\slash\!\!\!k}_B$. To identify consequences associated specifically with this constant, we examine the canonical uncertainty bounds and a separate quadratic entropy-production observable. Under the retained-mode and equilibrium matching assumptions described below, adjacent entropy levels satisfy
\begin{equation}\label{I3}\Delta\Sigma={\slash\!\!\!k}_B
\end{equation}
We revisit this entropy-quantization construction to specify the mode counting and distinguish its spectrum from that of the evolution generator. The predicted spacing is a conditional candidate signature of the reduced constant. It does not follow from the first-order factorization alone, and testing it requires a measurement prescription and sufficient experimental resolution. Published thermoelectric data for a bismuth telluride film illustrate the geometric construction \cite{Kurokawa2020}. At $ZT\simeq0.68$, equal displacements along the principal directions of normalized charge-heat coordinates have distances differing by a factor of approximately $2.12$. This example shows how transport measurements determine the geometry within the model. It does not test the entropy-level prediction.

\noindent Sections II and III introduce the flux observables and derive the compatible evolution generator. Section IV presents the spectral-distance construction, and Section V establishes the exact factorization and metric for coordinate-dependent coefficients. Sections VI and VII develop the Onsager Clifford–Dirac representation and compare the resulting distances. Section VIII treats the thermoelectric example. Section IX examines the separate entropy-production observable and its spectrum, while Section X discusses the results and the remaining physical questions.

% ============================================================
% MESOSCOPIC OBSERVABLES
% ============================================================

\section{Mesoscopic flux observables}\label{sec:flux_observables}

Let $\alpha_\mu$ denote thermodynamic coordinates and $X^\mu$ their conjugate forces. At the mesoscopic scale, the canonical commutation relations are \cite{Sonnino2024a, Sonnino2024b}
\begin{equation}
[\alpha^\mu,\widehat X^\nu]
=
i\varepsilon\delta^{\mu\nu},
\qquad
\varepsilon=
\frac{\slash\!\!\! k_B}{2}
\label{eq:CCR}
\end{equation}
In the coordinate representation,
\begin{equation}
\widehat X^\mu
=
-i\varepsilon\partial^\mu,
\qquad
\partial^\mu
=
\frac{\partial}{\partial\alpha_\mu}
\label{eq:X}
\end{equation}
Motivated by the general decomposition of Hermitian geometric tensors into symmetric and antisymmetric contributions \cite{Provost1980,QuantumGeometry2025}, we consider a positive Hermitian transport tensor of the form
\begin{equation}
\tau_{\mu\nu}(\alpha)
=
g_{\mu\nu}(\alpha)
+i f_{\mu\nu}(\alpha)
\end{equation}\label{eq:tau}
where
\begin{equation}
g_{\mu\nu}=g_{\nu\mu},
\qquad
f_{\mu\nu}=-f_{\nu\mu}
\label{eq:symmetry}
\end{equation}
and both tensors are real. We use the complex tensor to specify a Hermitian mesoscopic differential structure. Its entries do not describe a classical law that maps real forces into complex measured currents. Start with
\begin{equation}
{\widehat A}_\mu
=
\tau_{\mu\nu}\widehat X^\nu
\label{eq:A}
\end{equation}
When the coefficients vary, ${\widehat A}_\mu$ is generally not self-adjoint. We choose the Hermitian flux prescription
\begin{equation}
\widehat J_\mu=\frac12({\widehat A}_\mu+{\widehat A}_\mu^\dagger)
\label{eq:J}
\end{equation}
Using Eq.~(\ref{eq:X}), we obtain
\begin{equation}
\widehat J_\mu=-i\varepsilon\left(g_{\mu\nu}\partial^\nu+\frac12\partial^\nu g_{\mu\nu}\right)-\frac{\varepsilon}{2}\partial^\nu f_{\mu\nu}
\label{eq:Jfull}
\end{equation}
The first term is associated with
symmetric transport, whereas the
antisymmetric sector contributes
through
\begin{equation}
\widehat J_\mu^{(a)}
=
-\frac{\varepsilon}{2}
\partial^\nu f_{\mu\nu}
\label{eq:Ja}
\end{equation}
Varying antisymmetric coefficients therefore contribute to the observable flux. The Hermitian prescription supplies self-adjoint flux observables, but it remains a specific quantization choice. We do not identify the unprojected complex expression with a measured flux.

% ============================================================
% EVOLUTION GENERATOR
% ============================================================

\section{Evolution from flux observability}\label{sec:evolution}

We seek a self-adjoint operator $D$ whose coordinate commutators reproduce the mesoscopic flux observables:
\begin{equation}
i\varepsilon[D,\alpha^\mu]=\widehat J_\mu
\label{eq:Heisenberg}
\end{equation}
Reference~\cite{Sonnino2024a} motivates the mesoscopic evolution equation through the canonical description. Alongside the coordinate-force commutators, it proposes the formal analogy
\begin{equation}\label{ev1}
[t,\sigma]=i\varepsilon, \qquad \varepsilon=\frac{\slash\!\!\! k_B}{2}
\end{equation}
Time is a parameter here, so this relation is heuristic. It suggests the substitution $\sigma\rightarrow -i\varepsilon\partial_t$ and the evolution equation
\begin{equation}\label{ev1}
i\varepsilon\frac{\partial}{\partial t}|\psi\rangle=\widehat{\sigma}_{\mathrm{eff}}|\psi\rangle
\end{equation}
If the pure state satisfies the evolution equation~(\ref{ev1}), where $\widehat{\sigma}$ is a self-adjoint operator, an observable $A$, which does not explicitly depend on time, evolves according to the law \cite{Sakurai2020, Littlejohn2020}:
\begin{equation}\label{ev2}
\frac{d A}{d t}=\frac{2i}{\slash\!\!\! k_B}[\widehat{\sigma}_{\mathrm{eff}},A]
\end{equation}  
On the other hand, irreversible thermodynamics requires the evolution law for the following observable $\alpha_\mu$:
\begin{equation}
\frac{d\alpha_\mu}{dt}=J_\mu
\end{equation}
Compatibility with the Heisenberg-like evolution of the thermodynamic coordinate operators therefore imposes
\begin{equation}
\frac{i}{\varepsilon}[\widehat{\sigma}_{\mathrm{eff}},\alpha_\mu]=\widehat J_\mu
\end{equation}
Comparing this condition with $i\varepsilon[D,\alpha_\mu]=\widehat J_\mu$ identifies the minimal effective generator as
\begin{equation}
\widehat{\sigma}_{\mathrm{eff}} = \varepsilon^2 D
\end{equation}
Hence, the mesoscopic evolution equation becomes
\begin{equation}
i\frac{\partial}{\partial t}|\psi\rangle=\varepsilon D|\psi\rangle
\end{equation}
For the symmetric sector, define
\begin{equation}
D_0 =-\frac12\partial^\nu
(g_{\mu\nu}\partial^\mu)
\label{eq:D0}
\end{equation}
A direct calculation gives
\begin{equation}
i\varepsilon[D_0,\alpha_\mu]
=
-i\varepsilon
\left(
g_{\mu\nu}\partial^\nu
+
\frac12\partial^\nu g_{\mu\nu}
\right)
\label{eq:D0comm}
\end{equation}
To incorporate the antisymmetric sector, introduce
\begin{equation}
M
=
-\frac12
(\partial^\nu f_{\mu\nu})
\partial^\mu
\label{eq:M}
\end{equation}
Antisymmetry implies
\begin{equation}
\partial^\mu\partial^\nu
f_{\mu\nu}=0
\label{eq:divzero}
\end{equation}
so $M$ is formally skew-adjoint under suitable boundary conditions. Also,
\begin{equation}
\varepsilon[M,\alpha_\mu]=-\frac{\varepsilon}{2}\partial^\nu f_{\mu\nu}
\label{eq:Mcomm}
\end{equation}
The full generator is therefore
\begin{align}
\widehat{\sigma}_{\mathrm{eff}}&=\varepsilon^2 D=\varepsilon^2(D_0-iM)\\
&=-\frac{\varepsilon^2}{2}\partial^\nu\left[(g_{\mu\nu}-if_{\mu\nu})\partial^\mu\right]\noindent
\label{eq:D}
\end{align}
where the symmetry of $g_{\mu\nu}$ and the antisymmetry of $f_{\mu\nu}$ have been taken into account. Hence, the generator
can equivalently be written as
\begin{equation}
D=-\frac12\partial^\mu
\left[\tau_{\nu\mu}(\alpha)\partial^\nu\right],
\qquad
\tau_{\nu\mu}=g_{\mu\nu}-if_{\mu\nu}
\end{equation}
This representation makes explicit its
Hermitian divergence form and will be
used in the subsequent factorization. It satisfies
\begin{equation}
i\varepsilon[D,\alpha_\mu]=\widehat J_\mu
\label{eq:centralCommutator}
\end{equation}
This commutator links the evolution generator to the chosen flux observables. It fixes $D$ up to a real multiplication operator $U(\alpha)$, which commutes with the coordinates. We choose the minimal realization, $U=0$. The construction requires no affine connection from macroscopic TFT. Such a connection may enter nonlinear extensions, potentially subject to the Prigogine--Glansdorff Universal Criterion. The mesoscopic state evolves unitarily under Eq.~(17). This preserves its von Neumann entropy but does not rule out thermodynamic entropy production in a reduced or coarse-grained description. The latter requires its own observable and statistical interpretation. We keep these two notions of entropy separate.

\subsection{Physical interpretation of the mesoscopic evolution}

The physical evolution of the mesoscopic state obeys the equation
\begin{equation}
i\varepsilon\frac{\partial\psi}{\partial t}
=
\widetilde D\psi,
\qquad
\widetilde D=\varepsilon^2D
\label{eq:mesoscopic_schrodinger}
\end{equation}
where
\begin{equation}
D=-\frac12\partial^\mu
\left(\tau_{\nu\mu}(\alpha)\partial^\nu\right)
\end{equation}
is the normalized mesoscopic evolution operator. Here, $\psi(\alpha,t)$ represents the mesoscopic state in the space of thermodynamic coordinates $\alpha_\mu$, and $\tau_{\mu\nu}$ is the Hermitian transport tensor. Since $D$ is self-adjoint on the appropriate
Hilbert space, the formal solution is
\begin{equation}
\psi(t)
=
\exp(-i\varepsilon Dt)\psi(0)
\end{equation}
The evolution operator is unitary, implying
\begin{equation}
\|\psi(t)\|^2
=
\|\psi(0)\|^2
\end{equation}
If the mesoscopic probability density is
identified with
\begin{equation}
P(\alpha,t)=|\psi(\alpha,t)|^2
\end{equation}
the normalization condition
\begin{equation}
\int_\Omega P(\alpha,t)\,d^n\alpha=1
\end{equation}
is preserved throughout the evolution. The evolution equation therefore
describes the coherent, reversible, and
norm-preserving evolution of mesoscopic
probability amplitudes in thermodynamic-coordinate
space. Although the symmetric part of the
transport tensor determines the dissipative
quadratic form, its presence in the self-adjoint
generator does not, by itself, produce
irreversible relaxation. The emergence of
macroscopic dissipation requires additional
physical mechanisms, such as environmental
coupling or coarse-graining of unresolved
mesoscopic degrees of freedom.

% ============================================================
%SPECTRAL GEOMETRY OF MESOSCOPIC EVOLUTION
% ============================================================

\section{Spectral Geometry of Mesoscopic Evolution}\label{sec:spectral_geometry}

We now reconstruct distance from the evolution generator. In Connes' approach \cite{Connes1994}, a first-order operator acts on a Hilbert space together with an algebra of coordinate functions. Its commutators with those functions set a bound on their variation. For points $\alpha$ and $\beta$, the distance is
\begin{equation}
d(\alpha,\beta)=
\sup_{\substack{a=a^*\\
\|[\mathscr{D},a]\|\leq 1}}
|a(\alpha)-a(\beta)|
\label{eq:connes_distance}
\end{equation}
where the algebra acts by multiplication. Equation~(\ref{eq:connes_distance}) directs us to the first-order commutators. We therefore need an operator whose coefficients come from the mesoscopic generator. Recall the evolution equation
\begin{equation}
i\frac{\partial}{\partial t}|\psi\rangle=
\varepsilon D|\psi\rangle
\end{equation}
where $D$ is the self-adjoint second-order transport operator and $\widehat\sigma_{\mathrm{eff}}=\varepsilon^2D$. We seek a self-adjoint first-order operator $\mathscr D$ with
\begin{equation}
\mathscr{D}^{\,2}=D
\label{eq:dirac_square_goal}
\end{equation}
or an operator whose square recovers $D$ on a chosen sector of a larger Hilbert space. The aim is to factor the transport generator, rather than prescribe a separate metric. Positivity of the Hermitian transport tensor permits an exact first-order factorization. A graded enlargement makes the first-order operator self-adjoint and places $D$ in its scalar sector. For constant Onsager coefficients, a Clifford representation offers another realization. We will show that both give the same distance in that regime. The metric describes thermodynamic-coordinate space. Once reconstructed, it defines a Levi-Civita connection, although that connection need not coincide with the thermodynamic affine connection of TFT.

% ============================================================
% GENERAL DIRAC-TYPE FACTORIZATION AND SPECTRAL GEOMETRY}
% ============================================================
\section{General Dirac-Type Factorization and Spectral Geometry}\label{sec:general_dirac}

The factorization should retain the derivatives of the coefficients. We therefore consider a coordinate-dependent Hermitian tensor with both symmetric and antisymmetric parts. We first construct an exact self-adjoint graded factorization, then calculate the metric that its commutators determine.

\subsection{Quadratic mesoscopic evolution operator}

Let $\alpha^\mu$, $\mu=1,\ldots,n$, denote the
mesoscopic thermodynamic coordinates, and let
$\partial^\mu=\partial/\partial\alpha^\mu$.
The evolution generator obtained from the
observability of thermodynamic fluxes is
\begin{equation}
D=-\frac{1}{2}\partial^\mu\left[\tau_{\nu\mu}(\alpha)\partial^\nu\right]
\label{eq:general_evolution}
\end{equation}
where repeated indices are summed and
\begin{equation}
\tau_{\mu\nu}(\alpha)=g_{\mu\nu}(\alpha)+if_{\mu\nu}(\alpha)
\label{eq:general_transport}
\end{equation}
The real tensors $g_{\mu\nu}$ and $f_{\mu\nu}$ satisfy
\begin{equation}
g_{\mu\nu}=g_{\nu\mu},
\qquad
f_{\mu\nu}=-f_{\nu\mu}
\label{eq:general_symmetries}
\end{equation}
Hence, $\tau$ is Hermitian.
\begin{equation}
\tau_{\mu\nu}^\dag=\tau_{\mu\nu}\quad{\rm and}\quad \tau_{\nu\mu}^\dag=\tau_{\nu\mu}
\label{eq:general_hermiticity}
\end{equation}
We assume sufficient regularity and uniform positive definiteness: there is a constant $c>0$ such that
\begin{equation}
{z^\mu}^{*}\tau_{\mu\nu}(\alpha)z^\nu
\geq
c\sum_{\kappa=1}^{n}|z_\kappa|^2
\label{eq:general_positivity}
\end{equation}
for every $z\in\mathbb C^n$ and every $\alpha$. We work on $\mathbb R^n$ with Lebesgue measure $d^n\alpha$ and take the coefficients to have the regularity and boundedness required below. Other domains need suitable boundary conditions. For smooth compactly supported $\psi$, integration by parts gives
\begin{equation}
\langle\psi,D\psi\rangle
=
\frac{1}{2}
\int_{\mathbb R^n}d^n\alpha\,
(\partial^\mu\psi)^{*}
\tau_{\nu\mu}
(\partial^\nu\psi)
\label{eq:general_quadratic_form}
\end{equation}
Since $\tau$ is Hermitian and positive
definite, this quadratic form is real
and nonnegative:
\begin{equation}
\langle\psi,D\psi\rangle\geq0
\label{eq:general_D_positive}
\end{equation}
Closing this form defines a nonnegative self-adjoint realization of $D$. We use that positivity to construct the factorization.

\subsection{Exact first-order factorization}

Every positive-definite Hermitian matrix $\tau(\alpha)$ admits a factorization \cite{HornJohnson}
\begin{equation}
\tau(\alpha)=B^\dagger(\alpha)B(\alpha)
\label{eq:general_B_factorization}
\end{equation}
The positive Hermitian square root gives a convenient choice \cite{Higham}:
\begin{equation}
B(\alpha)=\tau^{1/2}(\alpha)
\label{eq:general_B_sqrt}
\end{equation}
In components, Eq.~(\ref{eq:general_B_factorization})
reads
\begin{equation}
\tau_{\nu\mu}
=
\sum_{a=1}^{n}
{B_{\mu}^a}^*B_{a\nu}
\label{eq:general_B_components}
\end{equation}
where ${B_{\mu}^a}^*=(B_{a\mu})^*$. We introduce the scalar Hilbert space
\begin{equation}
\mathcal H_0=L^2(\mathbb R^n)
\label{eq:general_H0}
\end{equation}
and the vector-valued Hilbert space
\begin{equation}
\mathcal H_1
=
L^2(\mathbb R^n,\mathbb C^n)
\label{eq:general_H1}
\end{equation}
Define the first-order differential operator
\begin{equation}
T:\mathcal H_0\longrightarrow\mathcal H_1
\label{eq:general_T_mapping}
\end{equation}
by
\begin{equation}
(T\psi)_=\frac{1}{\sqrt{2}}B_{a\mu}(\alpha)\partial^\mu\psi
\label{eq:general_T_definition}
\end{equation}
The adjoint of $T$, initially computed
on smooth compactly supported functions,
is
\begin{equation}
T^\dagger\Phi
=
-\frac{1}{\sqrt{2}}
\partial^\mu
\left(
{B_{\mu}}^{a*}\Phi_a
\right)
\label{eq:general_T_adjoint}
\end{equation}
The derivative in Eq.~(\ref{eq:general_T_adjoint}) acts on the full product ${B_{\mu}^a}^*\Phi_a$. Applying the adjoint to $T\psi$ gives
\begin{equation}
\begin{aligned}
T^\dagger T\psi
&=
-\frac{1}{2}
\partial^\mu
\left[
{B_{\mu}^a}^{*}
B_{a\nu}\partial^\nu\psi
\right]
\\[2mm]
&=
-\frac{1}{2}
\partial^\mu
\left[
\tau_{\nu\mu}
\partial^\nu\psi
\right]
\end{aligned}
\label{eq:general_TT_calculation}
\end{equation}
Comparison with Eq.~(\ref{eq:general_evolution})
immediately yields
\begin{equation}
D=T^\dagger T
\label{eq:general_exact_factorization}
\end{equation}
The factorization is exact. It includes the coefficient derivatives in both $g$ and $f$ and requires no constant-coefficient approximation.

\subsection{Self-adjoint graded Dirac-type operator}

The first-order operator $T$ maps scalars to vector-valued functions. It therefore does not act within the scalar Hilbert space alone. To obtain a self-adjoint first-order operator, use the graded construction \cite{Witten1982,Eckhardt2013} on
\begin{equation}
\mathcal H=\mathcal H_0\oplus\mathcal H_1
\label{eq:general_graded_space}
\end{equation}
and define the self-adjoint first-order operator
\begin{equation}
\mathscr{D}=
\begin{pmatrix}
0&T^\dagger\\
T&0
\end{pmatrix}
\label{eq:general_dirac_operator}
\end{equation}
A state in this space has the form
\begin{equation}
\Psi=
\begin{pmatrix}
\psi\\
\Phi
\end{pmatrix},
\qquad
\psi\in\mathcal H_0,
\quad
\Phi\in\mathcal H_1
\label{eq:general_graded_state}
\end{equation}
The action of $\mathscr{D}$ on $\Psi$ is
\begin{equation}
\mathscr{D}
\begin{pmatrix}
\psi\\
\Phi
\end{pmatrix}
=
\frac{1}{\sqrt{2}}
\begin{pmatrix}
-\partial^\mu({B_{\mu}^a}^{*}\Phi_a)\\[1mm]
B_{a\nu}\partial^\nu\psi
\end{pmatrix}
\label{eq:general_dirac_action}
\end{equation}
Take the closed realization of $T$. The block operator $\mathscr D$ is self-adjoint on
\begin{equation}
\operatorname{Dom}(\mathscr{D})
=
\operatorname{Dom}(T)
\oplus
\operatorname{Dom}(T^\dagger)
\label{eq:general_dirac_domain}
\end{equation}
The operator is naturally graded by
\begin{equation}
\Gamma=
\begin{pmatrix}
I_{\mathcal H_0}&0\\
0&-I_{\mathcal H_1}
\end{pmatrix}
\label{eq:general_grading}
\end{equation}
which satisfies
\begin{equation}
\Gamma^\dagger=\Gamma,
\qquad
\Gamma^2=I,
\qquad
\Gamma\mathscr{D}+\mathscr{D}\Gamma=0
\label{eq:general_grading_relations}
\end{equation}
Thus, $\mathscr{D}$ is an odd self-adjoint operator with respect to the grading. Squaring Eq.~(\ref{eq:general_dirac_operator}) gives
\begin{equation}
\mathscr{D}^2=
\begin{pmatrix}
T^\dagger T&0\\
0&TT^\dagger
\end{pmatrix}
\label{eq:general_dirac_square_intermediate}
\end{equation}
Using Eq.~(\ref{eq:general_exact_factorization}), we obtain
\begin{equation}
\mathscr{D}^2=
\begin{pmatrix}
D&0\\
0&TT^\dagger
\end{pmatrix}
\label{eq:general_dirac_square}
\end{equation}
Therefore, the original mesoscopic evolution generator is recovered exactly on the scalar sector:
\begin{equation}
\left.\mathscr{D}^2\right|_{\mathcal H_0}=D
\label{eq:general_scalar_recovery}
\end{equation}
The vector sector completes the self-adjoint representation. It adds no independent physical thermodynamic variables. Different factors $B$ satisfying Eq.~(\ref{eq:general_B_factorization}) give the same scalar operator $D$. The graded construction does not itself establish a Clifford algebra. We reserve the Clifford representation for the constant Onsager regime, where the additional algebraic relations hold. Here we use the exact graded factorization and its commutators to define distance.

\subsection{Role of the antisymmetric transport tensor}

Expanding Eq.~(\ref{eq:general_evolution}) separates the two transport contributions:
\begin{equation}
D=-\frac{1}{2}g_{\mu\nu}\partial^\mu\partial^\nu-\frac{1}{2}(\partial^\mu g_{\mu\nu})\partial^\nu+\frac{i}{2}(\partial^\mu f_{\mu\nu})\partial^\nu
\label{eq:general_D_expanded}
\end{equation}
because antisymmetry and commuting partial derivatives give $f_{\mu\nu}\partial^\mu\partial^\nu=0$. The remaining antisymmetric contribution is
\begin{equation}
D_f=\frac{i}{2}(\partial^\mu f_{\mu\nu})\partial^\nu
\label{eq:general_antisym_contribution}
\end{equation}
Constant $f$ contributes nothing to $D$. When $f$ varies, its derivatives contribute at first order. They leave the principal second-order part unchanged, a distinction that will also appear in the distance calculation.

\subsection{Spectral distance and reconstructed metric}

Let $\mathcal A$ be an algebra of bounded, sufficiently regular coordinate functions acting by multiplication on both graded sectors. For real $a(\alpha)$,
\begin{equation}
\begin{aligned}
[T,a]\psi
&=
T(a\psi)-aT\psi
\\
&=
\frac{1}{\sqrt{2}}
B_{a\mu}(\partial^\mu a)\psi
\end{aligned}
\label{eq:general_T_commutator}
\end{equation}
Thus, $[T,a]$ is a multiplicatio operator from scalar functions
to vector-valued functions. Since $a$ is real,
\begin{equation}
[T^\dagger,a]=-[T,a]^\dagger
\label{eq:general_adjoint_commutator}
\end{equation}
The commutator with the graded Dirac-type operator is therefore
\begin{equation}
[\mathscr{D},a]=
\begin{pmatrix}
0&-[T,a]^\dagger\\
[T,a]&0
\end{pmatrix}
\label{eq:general_graded_commutator}
\end{equation}
Its operator norm satisfies
\begin{equation}
\|[\mathscr{D},a]\|
=
\|[T,a]\|
\label{eq:general_commutator_norm_equal}
\end{equation}
Using the factorization
$\tau=B^\dagger B$, we obtain
\begin{equation}
\begin{aligned}
\|[T,a]\|^2
=
\frac{1}{2}
\operatorname*{ess\,sup}_{\alpha}
\Big[
(\partial^\mu a)
\tau_{\nu\mu}(\alpha)
(\partial^\nu a)
\Big]
\end{aligned}
\label{eq:general_commutator_tau}
\end{equation}
Substituting
$\tau_{\mu\nu}=g_{\mu\nu}+if_{\mu\nu}$
gives
\begin{equation}
\begin{aligned}
(\partial^\mu a)
\tau_{\nu\mu}
(\partial^\nu a)
={}&
g_{\mu\nu}
(\partial^\mu a)(\partial^\nu a)
\\
&-
if_{\mu\nu}
(\partial^\mu a)(\partial^\nu a)
\end{aligned}
\label{eq:general_tau_gradient}
\end{equation}
The last term vanishes identically as $f_{\mu\nu}(\partial^\mu a)(\partial^\nu a)=0$. It follows that
\begin{equation}
\|[\mathscr{D},a]\|
=
\operatorname*{ess\,sup}_{\alpha}
\sqrt{
\frac{1}{2}
g_{\mu\nu}(\alpha)
(\partial^\mu a)(\partial^\nu a)
}
\label{eq:general_commutator_norm}
\end{equation}
Only the symmetric tensor enters this norm. Define the Connes' \textit{spectral distance} between $\alpha$ and $\beta$ by
\begin{equation}
d_{\mathscr{D}}(\alpha,\beta)
=
\sup_{\substack{
a=a^*\in\mathcal A\\
\|[\mathscr{D},a]\|\leq1
}}
|a(\alpha)-a(\beta)|
\label{eq:general_connes_distance}
\end{equation}
The condition $a=a^*$ selects real multiplication functions. By Eq.~(\ref{eq:general_commutator_norm}), an admissible function satisfies
\begin{equation}
\frac{1}{2}
g_{\mu\nu}(\alpha)
(\partial^\mu a)(\partial^\nu a)
\leq1
\label{eq:general_lipschitz_constraint}
\end{equation}
almost everywhere. Comparing this with the standard Riemannian Lipschitz condition,
\begin{equation}
\|da\|_{G}^{2}=G_{\mu\nu}(\alpha)\partial^\mu a\partial^\nu a\leq 1
\label{eq:general_riemannian_norm}
\end{equation}
we identify the Connes' contravariant \textit{spectral metric} of the thermodynamic-coordinate space as
\begin{equation}
G_{\mu\nu}(\alpha)=\frac{1}{2}g_{\mu\nu}(\alpha)
\label{eq:general_contravariant_metric}
\end{equation}
Positivity of $\tau$ gives positivity of $g$ on real vectors. Its inverse therefore exists, and the covariant metric is
\begin{equation}
G^{\mu\nu}(\alpha)
=
2g^{\mu\nu}(\alpha)
\label{eq:general_covariant_metric}
\end{equation}
For a regular complete metric and a sufficiently rich function algebra, the spectral distance equals the geodesic distance:
\begin{equation}
d_{\mathscr{D}}(\alpha,\beta)=
\inf_{\gamma:\alpha\rightarrow\beta}
\int_0^1
\sqrt{
G^{\mu\nu}(\gamma(s))
\dot{\gamma}_{\mu}(s)
\dot{\gamma}_{\nu}(s)
}\,ds
\label{eq:general_geodesic_distance}
\end{equation}
The infimum is taken over piecewise smooth curves  $\alpha$ to $\beta$:
\begin{equation}
\gamma(0)=\alpha,
\qquad
\gamma(1)=\beta
\label{eq:general_curve_endpoints}
\end{equation}
A minimizing curve, when it exists, is a geodesic. Varying coefficients can produce curvature in thermodynamic-coordinate space. Notice that the resulting metric defines a Levi-Civita connection. It does not, however, derive the affine connection of TFT, which is generally not Levi-Civita.

\subsection{Recovery of the Onsager limit}

The constant-coefficient Onsager regime provides a useful consistency check. In this limit,
\begin{equation}
g_{\mu\nu}(\alpha)=L_{\mu\nu},
\qquad
f_{\mu\nu}=0
\label{eq:general_onsager_limit}
\end{equation}
where $L$ is a real, symmetric, positive-definite matrix. Equation~(\ref{eq:general_evolution}) reduces to
\begin{equation}
D=-\frac{1}{2}L_{\mu\nu}\partial^\mu\partial^\nu
\label{eq:general_onsager_D}
\end{equation}
The general metric formula becomes
\begin{equation}
G^{\mu\nu}=2L^{\mu\nu}
\label{eq:general_onsager_metric}
\end{equation}
in agreement with the result obtained from the Clifford-Dirac representation. The corresponding distance is
\begin{equation}
d_{\mathscr{D}}(\alpha,\beta)
=
\sqrt{
2(\alpha-\beta)_\mu
L^{\mu\nu}
(\alpha-\beta)_\nu
}
\label{eq:general_onsager_distance}
\end{equation}
In this special regime, the
constant positive-definite
Onsager matrix also permits
Hermitian matrices $\gamma^\mu$
satisfying
\begin{equation}
\{\gamma^\mu,\gamma^\nu\}
=
2L^{\mu\nu}I
\label{eq:general_onsager_clifford}
\end{equation}
The corresponding Clifford-Dirac operator,
\begin{equation}
\mathcal D=-\frac{i}{\sqrt{2}}L_{\mu\nu}\gamma^\mu\partial^\nu
\label{eq:general_onsager_clifford_D}
\end{equation}
satisfies
\begin{equation}
\mathcal D^2=DI
\label{eq:general_onsager_clifford_square}
\end{equation}
The graded and Clifford constructions give the same metric in the Onsager regime, although they act on different Hilbert spaces. For variable $g$, one can define a Clifford algebra at each point. Squaring a corresponding first-order operator introduces derivatives of the Clifford matrices and connection terms. Equality with $DI$ then needs further conditions. The graded factorization remains exact for the full variable Hermitian tensor.

\subsection{General result}

We can collect the factorization and metric formulas as follows:
\begin{align}
&D=-\frac{1}{2}\partial^\mu\left(\tau_{\nu\mu}\partial^\nu\right)\\
&\tau=B^\dagger B,\quad T=\frac{1}{\sqrt{2}}B\nabla,\quad D=T^\dagger T \nonumber
\end{align}
and
\begin{equation}
\mathscr{D}
=
\begin{pmatrix}
0&T^\dagger\\
T&0
\end{pmatrix},
\quad
\mathscr{D}^2
=
\begin{pmatrix}
D&0\\
0&TT^\dagger
\end{pmatrix}
\label{eq:general_summary_factorization}
\end{equation}
The associated contravariant spectral metric is
\begin{equation}
G^{\mu\nu}(\alpha)=2g^{\mu\nu}(\alpha)
\label{eq:general_summary_metric}
\end{equation}
The symmetric tensor sets the metric. Derivatives of the antisymmetric tensor can change the first-order dynamics without changing that metric. Both statements hold for variable coefficients.

% ============================================================
% EMERGENT CLIFFORD--DIRAC STRUCTURE
% ============================================================

\section{Emergent Clifford-Dirac Structure}\label{sec:clifford_dirac}

We now ask when the first-order structure admits a Clifford representation. Constant Onsager coefficients provide the simplest case: the transport tensor reduces to a real symmetric positive-definite matrix. The graded factorization remains available, but the constant matrix also fixes a Clifford algebra.

\subsection{Clifford algebra determined by the Onsager matrix}

In the linear Onsager regime, we have
\begin{equation}
g_{\mu\nu}=L_{\mu\nu},
\qquad
f_{\mu\nu}=0,
\qquad
L=L^{T}>0
\label{eq:clifford_onsager_conditions}
\end{equation}
The quadratic mesoscopic evolution generator therefore becomes
\begin{equation}
D=-\frac{1}{2}
L_{\mu\nu}\partial^\mu\partial^\nu
\label{eq:clifford_onsager_generator}
\end{equation}
Since $L$ is real, symmetric, and positive
definite, there exists an invertible real
matrix $E$ such that \cite{HornJohnson}:
\begin{equation}
L^{\mu\nu}=\sum_aE^{\mu}{}_{a}E^{\nu}{}_{a}
\label{eq:clifford_vielbein}
\end{equation}
Let $\Gamma^{a}$ denote Hermitian matrices satisfying the Euclidean Clifford relations
\begin{equation}
\{\Gamma^{a},\Gamma^{b}\}
=
2\delta^{ab}I
\label{eq:clifford_euclidean_algebra}
\end{equation}
We define
\begin{equation}
\gamma^{\mu}
=\sum_a E^{\mu}{}_{a}\Gamma^{a}
\label{eq:clifford_gamma_definition}
\end{equation}
It follows that
\begin{equation}
\begin{aligned}
\{\gamma^{\mu},\gamma^{\nu}\}
&=\sum_{a,b}
E^{\mu}{}_{a}E^{\nu}{}_{b}
\{\Gamma^{a},\Gamma^{b}\}
\\
&=
2\sum_aE^{\mu}{}_{a}E^{\nu}{}_{a}I
\end{aligned}
\label{eq:clifford_gamma_derivation}
\end{equation}
Therefore,
\begin{equation}
\{\gamma^{\mu},\gamma^{\nu}\}
=
2L^{\mu\nu}I
\label{eq:clifford_onsager_algebra}
\end{equation}
The Onsager matrix fixes the anticommutators in Eq.~(\ref{eq:clifford_onsager_algebra}). The matrices $\gamma^\mu$ thus introduce no independent transport parameters.

\subsection{Construction of the Clifford-Dirac operator}

We introduce the first-order operator
\begin{equation}
\mathcal D
=
-\frac{i}{\sqrt{2}}L_{\mu\nu}\gamma^{\mu}\partial^\nu
\label{eq:clifford_dirac_operator}
\end{equation}
It acts on the spinorial Hilbert space
\begin{equation}
\mathcal H_{\mathrm{sp}}
=
L^{2}(\mathbb R^{n},\mathbb C^{N}),
\qquad
N=2^{\lfloor n/2\rfloor}
\label{eq:clifford_spinor_space}
\end{equation}
The constant Hermitian matrices $\gamma^\mu$ give $\mathcal D$ a self-adjoint realization on $H^1(\mathbb R^n;\mathbb C^N)$ \cite{Thaller1992,ReedSimon1975}. Squaring Eq.~(\ref{eq:clifford_dirac_operator}) gives
\begin{equation}
\mathcal D^{2}=-\frac{1}{2}L_{\mu\nu}L_{\alpha\beta}\gamma^{\mu}\gamma^{\alpha}\partial^\nu\partial^\beta
\label{eq:clifford_square_first}
\end{equation}
Since the partial derivatives commute, only the symmetric part of the product $\gamma^\mu\gamma^\nu$ contributes:
\begin{equation}
\mathcal D^{2}=-\frac{1}{4}L_{\mu\nu}L_{\alpha\beta}\{\gamma^\mu,\gamma^\alpha\}\partial^\nu\partial^\beta
\label{eq:clifford_square_second}
\end{equation}
Using Eq.~(\ref{eq:clifford_onsager_algebra}), we find
\begin{equation}
\mathcal D^{2}
=
-\frac{1}{2}
L_{\mu\nu}
\partial^\mu\partial^\nu I
\label{eq:clifford_square_third}
\end{equation}
Comparison with
Eq.~(\ref{eq:clifford_onsager_generator}) gives the exact identity
\begin{equation}
\mathcal D^{2}=DI
\label{eq:clifford_exact_square}
\end{equation}
This is an exact Clifford square root for constant Onsager coefficients. The graded operator acts on scalar and vector functions; $\mathcal D$ acts on spinor functions. Their squares recover the same scalar generator in these different representations.

\subsection{Emergent spinorial representation}

For two independent thermodynamic
coordinates, the Clifford algebra
admits a representation in terms
of the Pauli matrices. Let
\begin{equation}
\Gamma^{1}=\sigma^{1},
\qquad
\Gamma^{2}=\sigma^{2}
\label{eq:clifford_pauli}
\end{equation}
where
\begin{equation}
\sigma^{1}
=
\begin{pmatrix}
0&1\\
1&0
\end{pmatrix},
\qquad
\sigma^{2}
=
\begin{pmatrix}
0&-i\\
i&0
\end{pmatrix}
\label{eq:clifford_pauli_matrices}
\end{equation}
These matrices satisfy
\begin{equation}
\{\sigma^a,\sigma^b\}
=
2\delta^{ab}I
\label{eq:clifford_pauli_algebra}
\end{equation}
The thermodynamic Clifford matrices
are therefore
\begin{equation}
\gamma^\mu
=
E^{\mu}{}_{1}\sigma^1
+
E^{\mu}{}_{2}\sigma^2
\label{eq:clifford_2d_gamma}
\end{equation}
The corresponding Dirac operator is
\begin{align}
\mathcal D=&-\frac{i}{\sqrt{2}}\Bigl(\left(L_{11}\gamma^1+L_{12}\gamma^2\right)\frac{\partial}{\partial\alpha_1}+\\
&\left(L_{12}\gamma^1+L_{22}\gamma^2\right)\frac{\partial}{\partial\alpha_2}\Bigr)\nonumber
\label{eq:clifford_2d_dirac}
\end{align}
It acts on a two-component function
\begin{equation}
\Psi(\alpha)
=
\begin{pmatrix}
\psi_1(\alpha)\\
\psi_2(\alpha)
\end{pmatrix}
\label{eq:clifford_two_component_spinor}
\end{equation}
The two components implement the Clifford algebra. They imply neither independent physical thermodynamic variables nor physical spin-$1/2$ particles. We use ``spinorial'' to describe this algebraic representation.

\subsection{Spectral branches and mesoscopic evolution}

To examine the spectrum of the first-order operator, consider a Fourier mode
\begin{equation}
\Psi(\alpha)
=
u(k)e^{ik^\mu\alpha_\mu}
\label{eq:clifford_fourier_mode}
\end{equation}
where $u(k)$ is a constant spinor. The corresponding matrix symbol is
\begin{equation}
\mathcal D(k)
=
\frac{1}{\sqrt{2}}
L_{\mu\nu}\gamma^\mu k^\nu
\label{eq:clifford_dirac_symbol}
\end{equation}
The Clifford relations imply
\begin{equation}
\mathcal D(k)^2
=
\frac{1}{2}
L_{\mu\nu}k^\mu k^\nu I
\label{eq:clifford_symbol_square}
\end{equation}
For the two-dimensional representation, the eigenvalues of the first-order symbol are
\begin{equation}
\lambda_\pm(k)
=
\pm
\sqrt{
\frac{1}{2}
L_{\mu\nu}k^\mu k^\nu
}
\label{eq:clifford_symbol_eigenvalues}
\end{equation}
The two signs belong to the first-order factorization. They do not identify distinct thermodynamic excitation branches, since the quadratic operator still governs evolution:
\begin{equation}
i\frac{\partial\Psi}{\partial t}
=
\varepsilon\mathcal D^2\Psi
=
\varepsilon DI\Psi
\label{eq:clifford_mesoscopic_evolution}
\end{equation}
For a Fourier mode, the associated
dispersion relation is
\begin{equation}
\omega(k)
=
\frac{\varepsilon}{2}
L_{\mu\nu}k^\mu k^\nu
\label{eq:clifford_dispersion}
\end{equation}
Thus, the Clifford-Dirac representation preserves the quadratic mesoscopic evolution law.

\subsection{Relation to the general Dirac-type factorization}

Constant Onsager coefficients therefore fix a Clifford algebra. The more general graded identity,
\begin{equation}
\mathscr{D}
=
\begin{pmatrix}
0&T^\dagger\\
T&0
\end{pmatrix},
\qquad
D=T^\dagger T
\label{eq:clifford_general_comparison}
\end{equation}
remains valid for positive
Hermitian transport tensors
with coordinate-dependent
symmetric and antisymmetric
components. By contrast, the exact identity
\begin{equation}
\mathcal D^2=DI
\label{eq:clifford_special_identity}
\end{equation}
holds here for constant symmetric coefficients. If those coefficients vary, the Clifford matrices vary too. Their derivatives and any connection terms enter the square of the first-order operator. A pointwise Clifford algebra alone does not give an exact square root of the full generator. The two constructions serve different purposes. The graded factorization works for the positive Hermitian tensor, including its variable antisymmetric part. The Clifford representation realizes the first-order structure under the constant Onsager assumptions.

% ============================================================
% SPECTRAL DISTANCE AND THERMODYNAMIC METRIC
% ============================================================
\section{Spectral Distance and Thermodynamic Metric}\label{sec:spectral_metric}

The operators $\mathscr D$ and $\mathcal D$ recover the same generator in the Onsager regime. Do they also give the same distance? We now compare their commutator norms using the same algebra of coordinate functions.

\subsection{Spectral distance from the Clifford-Dirac operator}

Consider the constant-coefficient
Onsager regime, where
\begin{equation}
g_{\mu\nu}=L_{\mu\nu},
\qquad
f_{\mu\nu}=0,
\qquad
L=L^T>0
\end{equation}
The mesoscopic evolution generator is
\begin{equation}
D=-\frac12 L_{\mu\nu}\partial^\mu\partial^\nu
\end{equation}
The corresponding Clifford-Dirac operator is
\begin{equation}
\mathcal D=-\frac{i}{\sqrt2}L_{\mu\nu}\gamma^\mu\partial^\nu
\end{equation}
where
\begin{equation}
\{\gamma^\mu,\gamma^\nu\}=2L^{\mu\nu}I
\end{equation}
Let a real, sufficiently regular function $a(\alpha)$ act by multiplication on spinor functions. Its commutator with $\mathcal D$ is
\begin{equation}
\begin{aligned}
[\mathcal D,a]\Psi
&=\mathcal D(a\Psi)-a\mathcal D\Psi
\\
&=-\frac{i}{\sqrt2}L_{\mu\nu}\gamma^\mu(\partial^\nu a)\Psi
\end{aligned}
\end{equation}
Hence,
\begin{equation}
[\mathcal D,a]
=-\frac{i}{\sqrt2}L_{\mu\nu}\gamma^\mu\partial^\nu a
\label{eq:metric_clifford_commutator}
\end{equation}
The Clifford relations determine the norm. For real $a$,
\begin{equation}
\begin{aligned}
\left(
L_{\mu\nu}\gamma^\mu\partial^\nu a
\right)^2
&=
\frac{1}{2}L_{\mu\nu}L_{\alpha\beta}
\{\gamma^\mu,\gamma^\alpha\}
(\partial^\nu a)(\partial^\beta a)
\\
&=L_{\mu\nu}
(\partial^\mu a)(\partial^\nu a)I
\end{aligned}
\end{equation}
Therefore,
\begin{equation}
\|[\mathcal D,a]\|
=
\operatorname*{ess\,sup}_{\alpha}
\sqrt{
\frac12
L_{\mu\nu}
(\partial^\mu a)(\partial^\nu a)
}
\label{eq:metric_clifford_norm}
\end{equation}
Connes' spectral distance between
two thermodynamic-coordinate
points $\alpha$ and $\beta$ is
\begin{equation}
d_{\mathcal D}(\alpha,\beta)
=
\sup_{\substack{
a=a^*\\
\|[\mathcal D,a]\|\leq1
}}
|a(\alpha)-a(\beta)|
\label{eq:metric_connes_distance}
\end{equation}
The constraint in this expression is equivalent to
\begin{equation}
\frac12
L_{\mu\nu}
(\partial^\mu a)(\partial^\nu a)
\leq1
\end{equation}
Comparing this condition with the Riemannian Lipschitz norm,
\begin{equation}
G_{\mu\nu}
(\partial^\mu a)(\partial^\nu a)
\leq1
\end{equation}
we identify
\begin{equation}
G_{\mu\nu}
=
\frac12L_{\mu\nu}
\label{eq:metric_contravariant}
\end{equation}
The corresponding covariant metric is
\begin{equation}
G^{\mu\nu}
=
2L^{\mu\nu}
\label{eq:metric_covariant}
\end{equation}
The Onsager coefficients thus determine the metric without an additional geometric tensor.

\subsection{Explicit spectral distance}

For constant $L_{\mu\nu}$, the minimizing curves (the geodesics) are the straight line
\begin{equation}\label{geoOns}
\gamma_\mu(s)=\alpha_\mu+s(\beta_\mu-\alpha_\mu)
\end{equation}
This yields the following expression for the distance
\begin{equation}
d_{\mathcal D}(\alpha,\beta)
=
\sqrt{
2(\alpha-\beta)_\mu
L^{\mu\nu}
(\alpha-\beta)_\nu
}
\label{eq:metric_explicit_distance}
\end{equation}
Positivity of $L$ ensures a positive-definite distance. For equal coordinate displacements along transport eigenvectors, larger transport eigenvalues give shorter distances. The Onsager matrix therefore fixes the metric anisotropy. The metric is flat. Its constant entries give a vanishing Levi-Civita connection in these affine coordinates and zero Riemann curvature, even when the principal distances differ. These distances concern thermodynamic coordinates, not lengths in physical space.

\subsection{Equivalence with the general graded construction}

To compare the two representations, recall the norm from the graded factorization $D=T^\dagger T$:
\begin{equation}
\|[\mathscr{D},a]\|
=
\operatorname*{ess\,sup}_{\alpha}
\sqrt{
\frac12 g_{\mu\nu}(\alpha)
(\partial^\mu a)(\partial^\nu a)
}
\end{equation}
In the constant-coefficient Onsager regime,
$g_{\mu\nu}=L_{\mu\nu}$ and $f_{\mu\nu}=0$.
Hence,
\begin{equation}
\|[\mathscr{D},a]\|
=
\operatorname*{ess\,sup}_{\alpha}
\sqrt{
\frac12 L_{\mu\nu}
(\partial^\mu a)(\partial^\nu a)
}
\end{equation}
Comparison with the commutator norm obtained
above for the Clifford--Dirac operator
$\mathcal D$ gives
\begin{equation}
\|[\mathscr{D},a]\|
=
\|[\mathcal D,a]\|
\end{equation}
Thus, when the same algebra of thermodynamic-coordinate
functions is used, both first-order constructions
determine the same spectral distance:
\begin{equation}
d_{\mathscr{D}}(\alpha,\beta)
=
d_{\mathcal D}(\alpha,\beta)
\end{equation}
The two auxiliary Hilbert spaces differ, but their commutator norms agree. They therefore define the same distance with the same function algebra. Both give
\begin{equation}
G^{\mu\nu}=2L^{\mu\nu}
\end{equation}

\subsection{Coordinate-dependent transport coefficients}

We now return to the general case in which the symmetric transport coefficients depend on the thermodynamic coordinates:
\begin{equation}
g_{\mu\nu}=g_{\mu\nu}(\alpha)
\end{equation}
The commutator norm obtained from the graded operator is
\begin{equation}
\|[\mathscr{D},a]\|
=
\operatorname*{ess\,sup}_{\alpha}
\sqrt{
\frac12
g_{\mu\nu}(\alpha)
(\partial^\mu a)(\partial^\nu a)
}
\end{equation}
Therefore, the local contravariant metric is
\begin{equation}
G_{\mu\nu}(\alpha)
=
\frac12g_{\mu\nu}(\alpha)
\label{eq:metric_general_contravariant}
\end{equation}
and the covariant metric is
\begin{equation}
G^{\mu\nu}(\alpha)
=
2g^{\mu\nu}(\alpha)
\label{eq:metric_general_covariant}
\end{equation}
Under appropriate regularity
and completeness assumptions,
the spectral distance agrees
with the geodesic distance
associated with this metric:
\begin{equation}
d_{\mathscr{D}}(\alpha,\beta)
=
\inf_{\gamma:\alpha\rightarrow\beta}
\int_0^1
\sqrt{
G^{\mu\nu}(\gamma(s))
\dot{\gamma}_{\mu}(s)
\dot{\gamma}_{\nu}(s)
}\,ds
\label{eq:metric_general_geodesic}
\end{equation}
Here the infimum runs over piecewise smooth paths from $\alpha$ to $\beta$. A varying metric can have nonzero curvature. The antisymmetric tensor still drops out because it contracts to zero with the symmetric gradient product. Its derivatives can nevertheless enter the first-order terms in $D$. The full tensor affects evolution; its symmetric part fixes the metric.

\subsection{Relation to Thermodynamic Field Theory}

The metric and the affine connection of TFT \cite{Sonnino2009,Sonnino2016,Sonnino2022,Sonnino2002,Sonnino2019,Sonnino2001} encode different structures. TFT's connection is generally not the Levi-Civita connection of the reconstructed metric. Our calculation determines
\begin{equation}
G^{\mu\nu}(\alpha)
=
2g^{\mu\nu}(\alpha)
\end{equation}
but leaves the thermodynamic affine connection undetermined. The metric defines its own Levi-Civita connection, which should not be identified with the TFT connection. Encoding the latter through additional operators would require further physical input. The present factorization supplies a distance and a metric directly from transport.

% ============================================================
% THERMODYNAMIC REALIZATION
%============================================================

\section{Thermodynamic Realization}\label{sec:thermodynamic_realization}

\begin{figure}[t]
\centering
\includegraphics[width=0.8\columnwidth]{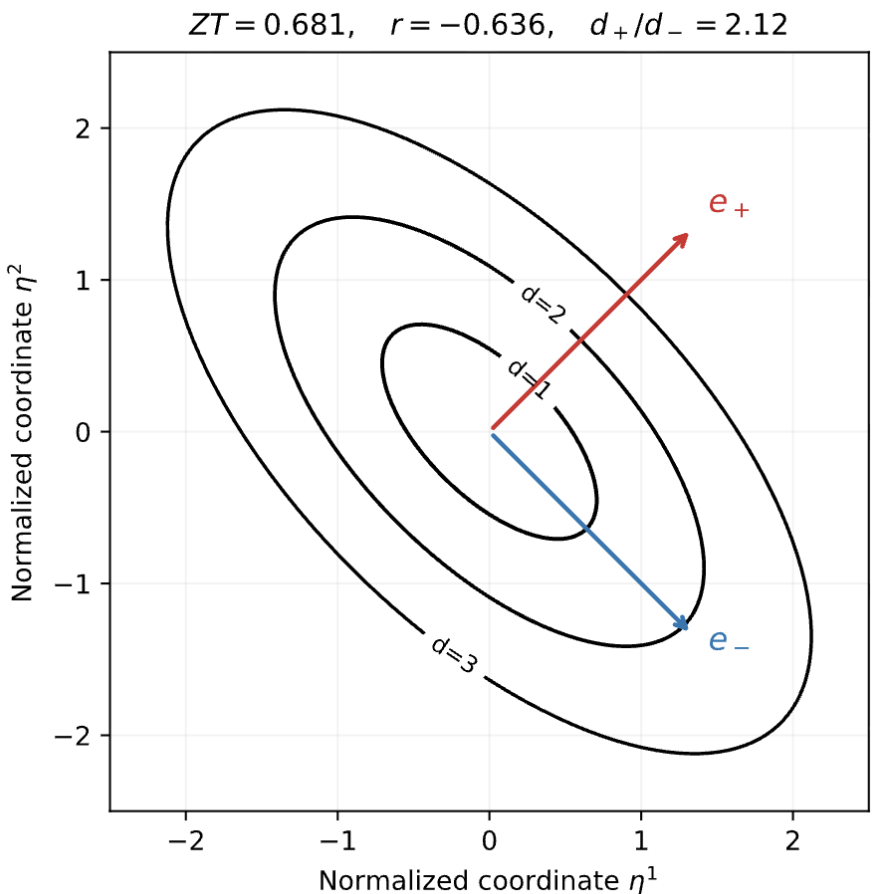}
\caption{
Contours of constant spectral distance
$d(0,\eta)$ in normalized thermodynamic-coordinate
space for the $\mathrm{Bi_2Te_3}$ thin film
considered in Table~I.
The reconstructed metric is
$G_{(\eta)}=2\widetilde L^{-1}$,
with $ZT\simeq0.681$ and $r\simeq-0.636$.
The principal directions are
$\mathbf e_\pm=(1,\pm1)/\sqrt{2}$. The elliptical contours illustrate the
anisotropy induced by thermoelectric coupling.
For equal normalized coordinate displacements
along the principal directions, the ratio
of spectral distances is approximately $2.12$.
The distances refer to thermodynamic-coordinate
space, not physical space.}
\label{fig:spectral_contours}
\end{figure}

We illustrate the calculation with a nanocrystalline $\mathrm{Bi_2Te_3}$ film deposited on glass by magnetron sputtering and then annealed \cite{Kurokawa2020}. Electrical and thermal transport couple strongly near room temperature in this material. Set $J_1=j_e$ and $J_2=j_q$, the electrical and heat current densities. Along the transport direction $x$, use the forces \cite{deGroot1984}
\begin{equation}
X^1=\frac{E_{\mathrm{ec}}}{T},
\qquad
X^2=\frac{d}{dx}\left(\frac{1}{T}\right)
\end{equation}
where $E_{\mathrm{ec}}=-dV_{\mathrm{ec}}/dx$ is the field
derived from the electrochemical potential per unit charge.
These choices give the local entropy-production rate per unit volume
\begin{equation}
\sigma_s=j_eX^1+j_qX^2
\end{equation}
This force convention applies when the second flux is
the heat current density. With this choice of flux convention we get \cite{Glansdorff1971}:
\begin{equation}
\frac{d\alpha_1}{dt}=j_e,
\qquad
\frac{d\alpha_2}{dt}=j_q
\end{equation}
Integrating these current densities defines $\alpha^1$ as transferred charge per unit area and $\alpha^2$ as transferred heat per unit area, relative to a reference state. Their SI units are $\mathrm{C\,m^{-2}}$ and $\mathrm{J\,m^{-2}}$. Table~\ref{tab:bi2te3_data} lists the reported room-temperature parameters.
\begin{table}[t]
\caption{
Room-temperature thermoelectric
parameters of a nanocrystalline
$\mathrm{Bi_2Te_3}$ thin film reported in
Ref.~\cite{Kurokawa2020}. The electrical
conductivity and Seebeck coefficient were
measured, whereas the thermal conductivity
was estimated using a transport model.
}
\label{tab:bi2te3_data}
\begin{ruledtabular}
\begin{tabular}{lc}
Quantity & Value\\
\hline
Temperature $T$ & $300\ \mathrm{K}$\\
Electrical conductivity $\sigma$
& $1.025\times10^{5}\ \mathrm{S\,m^{-1}}$\\
Seebeck coefficient $S$
& $-163\ \mu\mathrm{V\,K^{-1}}$\\
Thermal conductivity $\kappa$
& $1.20\ \mathrm{W\,m^{-1}K^{-1}}$\\
Reported $ZT$ & $\simeq0.70$
\end{tabular}
\end{ruledtabular}
\end{table}
\noindent The negative Seebeck coefficient indicates electron-dominated transport. We use the tabulated values without adjustment to calculate
\begin{equation}
ZT=\frac{\sigma S^2T}{\kappa}
\label{eq:bi2te3_zt}
\end{equation}
Substituting the experimental parameters gives
\begin{equation}
ZT
=
\frac{
(1.025\times 10^5)
(1.63\times10^{-4})^2
(300)
}{1.20}
\simeq0.681\nonumber 
\label{eq:bi2te3_zt_numerical}
\end{equation}
The calculated $ZT$ agrees with the reported estimate $ZT\simeq0.70$ at the precision of the supplied parameters. The coupling coefficient is
\begin{equation}
r=\operatorname{sgn}(S)
\sqrt{\frac{ZT}{1+ZT}}
\label{eq:bi2te3_r}
\end{equation}
For the present sample,
\begin{equation}
r\simeq-0.636
\label{eq:bi2te3_r_numerical}
\end{equation}
Introducing
\begin{equation}
S=\operatorname{diag}
\left(\sqrt{L_{11}},\sqrt{L_{22}}\right)
\end{equation}
Onsager's matrix can be written as $L=S\widetilde L S$, where
\begin{equation}
\widetilde L=
\begin{pmatrix}
1&r\\
r&1
\end{pmatrix}
\end{equation}
Since $\eta=S^{-1}\alpha$, the covariant spectral metric transforms according to
\begin{equation}
G_{(\eta)}
=S^T G_{(\alpha)}S
=2\widetilde L^{-1}
\end{equation}
The magnitude of $r$ indicates appreciable electrical--thermal coupling. To compare the two channels, use the normalized coordinates
\begin{equation}
\eta^\mu=\frac{\alpha^\mu}{\sqrt{L_{\mu\mu}}},\qquad\text{no sum over }\mu
\label{eq:bi2te3_coordinates}
\end{equation}
introduced previously. In these coordinates, the
dimensionless normalized Onsager matrix is
\begin{equation}
\widetilde L
=
\begin{pmatrix}
1&r\\
r&1
\end{pmatrix}
\label{eq:bi2te3_normalized_onsager}
\end{equation}
For the measured thermoelectric coefficients, this becomes
\begin{equation}
\widetilde L
\simeq
\begin{pmatrix}
1&-0.636\\
-0.636&1
\end{pmatrix}
\label{eq:bi2te3_onsager_numerical}
\end{equation}
The spectral metric reconstructed from the Onsager matrix is
\begin{equation}
G_{(\eta)}=2\widetilde L^{-1}
\label{eq:bi2te3_metric_definition}
\end{equation}
Therefore,
\begin{equation}
G_{(\eta)}
=
\frac{2}{1-r^2}
\begin{pmatrix}
1&-r\\
-r&1
\end{pmatrix}
\label{eq:bi2te3_metric_formula}
\end{equation}
Using $r=-0.636$, we obtain
\begin{equation}
G_{(\eta)}
\simeq
\begin{pmatrix}
3.36&2.14\\
2.14&3.36
\end{pmatrix}
\label{eq:bi2te3_metric_numerical}
\end{equation}
The transport inputs therefore give a numerical metric. Its principal directions are
\begin{equation}
e_+
=
\frac{1}{\sqrt2}
\begin{pmatrix}
1\\
1
\end{pmatrix},
\qquad
e_-
=
\frac{1}{\sqrt2}
\begin{pmatrix}
1\\
-1
\end{pmatrix}
\label{eq:bi2te3_principal_directions}
\end{equation}
The corresponding metric eigenvalues are
\begin{equation}
G_+=\frac{2}{1+r},
\qquad
G_-=\frac{2}{1-r}
\label{eq:bi2te3_eigenvalues}
\end{equation}
Numerically,
\begin{equation}
G_+\simeq5.50,
\qquad
G_-\simeq1.22
\label{eq:bi2te3_eigenvalues_numerical}
\end{equation}
Consider now two thermodynamic displacements of equal magnitude $\delta$ in the normalized coordinate space:
\begin{equation}
\Delta\eta_+=\delta e_+,
\qquad
\Delta\eta_-=\delta e_-
\label{eq:bi2te3_displacements}
\end{equation}
Their spectral distances are
\begin{equation}
d_+=\delta\sqrt{G_+},
\qquad
d_-=\delta\sqrt{G_-}
\label{eq:bi2te3_distances}
\end{equation}
For a unit normalized displacement, $\delta=1$, we find
\begin{equation}
d_+\simeq2.35,
\qquad
d_-\simeq1.11
\label{eq:bi2te3_distances_numerical}
\end{equation}
The ratio is therefore
\begin{equation}
\frac{d_+}{d_-}
=
\sqrt{\frac{1-r}{1+r}}
\simeq2.12
\label{eq:bi2te3_anisotropy}
\end{equation}
Figure~\ref{fig:spectral_contours} shows the constant-distance ellipses. Their axes follow the eigenvectors of the metric. The distance $d$ runs from the reference point $\eta=0$ to $\eta=(\eta^1,\eta^2)^T$:
\begin{equation}
d(0,\eta)=\sqrt{\eta^T G_{(\eta)}\eta},
\qquad
G_{(\eta)}=2\widetilde L^{-1}
\end{equation}
For the thermoelectric sample considered here,
\begin{equation}
d(0,\eta)\!\simeq\!\Big[3.36(\eta^1)^2\!+\!4.28\eta^1\eta^2\!+\!3.36(\eta^2)^2\Big]^{1/2}
\end{equation}
Each ellipse connects points at the same distance from the reference state. The coordinates describe thermodynamic transfers, not physical positions in the film.
\noindent Equal normalized displacements along the two principal directions therefore differ in distance by a factor of $2.12$. For comparison, setting $S=0$ removes the coupling, reduces the normalized Onsager matrix to the identity, and gives

\begin{equation}
G_{(\eta)}=2I
\label{eq:bi2te3_uncoupled_metric}
\end{equation}
and both principal directions
would have the same distance
for equal normalized displacements:
\begin{equation}
d_+=d_-=\sqrt2\,\delta
\label{eq:bi2te3_uncoupled_distance}
\end{equation}
The coupling makes the normalized metric anisotropic. This is anisotropy in thermodynamic-coordinate space, not in the crystal's physical space. Normalization matters because the original transport channels have different units; the coupling $r$ is dimensionless. Increasing $ZT$ increases $|r|$ and the ratio of principal distances. For this film, $ZT\simeq0.68$ gives $|r|\simeq0.64$ and a ratio of approximately $2.12$. These values come from the characterized transport coefficients alone. The calculation reconstructs a geometry from material data. It does not independently predict thermoelectric performance or measure spectral distances experimentally.

% ============================================================
%ENTROPY-PRODUCTION QUANTIZATION
% ============================================================

\section{Entropy Production Quantization}
\label{sec:entropy_quantization}

The evolution generator and the entropy-production observable describe
different aspects of the mesoscopic framework. Here we apply the
canonical quadratic entropy-production formalism to the thermoelectric
example. We first specify the retained-mode convention, so that the
number of thermodynamic forces, the cutoff and the equilibrium
contribution are treated consistently throughout.
We use $\Sigma$ for the entropy-valued observable corresponding to
$\Delta_I S$ in the correspondence calculation shown in \cite{Sonnino2024a}, not for an entropy-production rate.

\subsection{Canonical mesoscopic variables}

Let $n$ denote the number of independent thermodynamic forces and let
$\mathcal K_M$ contain $M$ independent retained mode labels $\mathbf K$
per channel. Thus the model contains $nM$ canonical pairs. The integer
$M=|\mathcal K_M|$ counts independent modes, rather than a maximum
wavevector magnitude. If a cutoff is specified in wavevector space,
$M$ is obtained by counting the independent modes below it, including
any reality constraints. For $\mu=1,\ldots,n$ and $\mathbf K\in\mathcal K_M$, impose
\begin{equation}
 [\xi_{\mu,\mathbf K},Y^\nu_{\mathbf K'}]
 =i\varepsilon\delta_\mu{}^\nu\delta_{\mathbf K\mathbf K'}I
 \label{eq:entropy_commutator}
\end{equation}
with
\begin{equation}
 \varepsilon=\frac{\slash\!\!\! k_B}{2}
 \label{eq:entropy_epsilon}
\end{equation}
We use $\slash\!\!\! k_B$ consistently for the mesoscopic scale denoted by a
slashed $k_B$ in the original convention. It is distinct from the
Boltzmann constant $k_B$. The remaining commutators vanish:
\begin{equation}
 \begin{aligned}
 [\xi_{\mu,\mathbf K},\xi_{\nu,\mathbf K'}]&=0,\\
 [Y^\mu_{\mathbf K},Y^\nu_{\mathbf K'}]&=0
 \end{aligned}
 \label{eq:entropy_other_commutators}
\end{equation}
Let $\ell=(\ell_{\mu\nu})$ be constant, real, symmetric and positive
definite, with $\ell^{\mu\nu}=(\ell^{-1})_{\mu\nu}$.
For each retained mode, define
\begin{equation}
 \widehat\Sigma_{\mathbf K}
 =\left(
 \bm\xi_{\mathbf K}^{T}\ell^{-1}\bm\xi_{\mathbf K}
 +\bm Y_{\mathbf K}^{T}\ell\bm Y_{\mathbf K}\right)
 \label{eq:entropy_quadratic_operator}
\end{equation}
where the columns have components $\xi_{\mu,\mathbf K}$ and
$Y^\mu_{\mathbf K}$. The full truncated observable is
\begin{equation}
 \widehat\Sigma_M
 =\sum_{\mathbf K\in\mathcal K_M}\widehat\Sigma_{\mathbf K}\geq0
 \label{eq:entropy_positive}
\end{equation}
Both members of every canonical pair enter this quadratic observable.
It is distinct from the first-order operator used in the distance
construction and is not obtained merely by squaring that operator.

\subsection{Reduction to independent harmonic oscillators}

The matrix $\ell$ has unique symmetric positive definite square roots.
For an orthogonal diagonalization $\ell=O\Lambda O^T$, they are
\begin{equation}
 \begin{aligned}
 \ell^{\pm1/2}&=O\operatorname{diag}
       (\lambda_1^{\pm1/2},\ldots,\lambda_n^{\pm1/2})O^T,\\
 \ell^{-1/2}&=(\ell^{1/2})^{-1}
 \end{aligned}
 \label{eq:entropy_matrix_roots}
\end{equation}
These are matrix functions, not entrywise square roots. At every
retained $\mathbf K$, set
\begin{equation}
 \bm q_{\mathbf K}=\ell^{-1/2}\bm\xi_{\mathbf K},
 \qquad
 \bm p_{\mathbf K}=\ell^{1/2}\bm Y_{\mathbf K}
 \label{eq:entropy_q}
\end{equation}
The inverse transformation is
\begin{equation}
 \bm\xi_{\mathbf K}=\ell^{1/2}\bm q_{\mathbf K},
 \qquad
 \bm Y_{\mathbf K}=\ell^{-1/2}\bm p_{\mathbf K}
 \label{eq:entropy_p}
\end{equation}
The entries of $\bm q_{\mathbf K},\bm p_{\mathbf K}$ are written as
$q_{\mathbf K}^{(\mu)},p_{\mathbf K}^{(\mu)}$, where the parenthesized
index labels a normalized canonical mode. The two transformation
matrices are inverse transposes. In particular,
\begin{equation}
 \ell^{-1/2}(\ell^{1/2})^T=I_n,
\end{equation}
so the normalized commutators are
\begin{equation}
 [q_{\mathbf K}^{(\mu)},p_{\mathbf K'}^{(\nu)}]
 =i\varepsilon\delta_{\mu\nu}\delta_{\mathbf K\mathbf K'}I
 \label{eq:entropy_canonical_normalized}
\end{equation}
The $q$--$q$ and $p$--$p$ commutators vanish. Substitution gives
\begin{equation}
 \widehat\Sigma_{\mathbf K}
 =\sum_{\mu=1}^{n}
 \left[(q_{\mathbf K}^{(\mu)})^2
      +(p_{\mathbf K}^{(\mu)})^2\right]
 \label{eq:entropy_oscillator_form}
\end{equation}
Thus each retained label contributes $n$ oscillators, and the full
observable contains $nM$ oscillators. The reciprocal coefficients
$\ell^{-1}$ and $\ell$ cancel from the normalized expression.
We use the standard irreducible canonical representation; equivalently,
the quadratic expressions can first be defined on the Schwartz space
in $L^2(\mathbb R^{nM})$ and then given their self-adjoint oscillator
closures.

\subsection{Creation operators and the equilibrium cutoff}

Define at each channel and retained label
\begin{equation}
 \begin{aligned}
 a_{\mathbf K}^{(\mu)}
 &=\frac{q_{\mathbf K}^{(\mu)}+ip_{\mathbf K}^{(\mu)}}
         {\sqrt{2\varepsilon}},\\
 {a_{\mathbf K}^{(\mu)}}^\dagger
 &=\frac{q_{\mathbf K}^{(\mu)}-ip_{\mathbf K}^{(\mu)}}
         {\sqrt{2\varepsilon}},\\
 N_{\mathbf K}^{(\mu)}
 &={a_{\mathbf K}^{(\mu)}}^\dagger a_{\mathbf K}^{(\mu)}
 \end{aligned}
 \label{eq:entropy_annihilation}
\end{equation}
Their commutation relation is
\begin{equation}
 [a_{\mathbf K}^{(\mu)},{a_{\mathbf K'}^{(\nu)}}^\dagger]
 =\delta_{\mu\nu}\delta_{\mathbf K\mathbf K'}I
 \label{eq:entropy_ladder_commutator}
\end{equation}
with annihilation operators mutually commuting and creation operators
mutually commuting. The single-oscillator identity \cite{epq-tong} is
\begin{equation}
 \begin{aligned}
 &\left[(q_{\mathbf K}^{(\mu)})^2
              +(p_{\mathbf K}^{(\mu)})^2\right]\\
 &\hspace{1em}=2\varepsilon
       \left(N_{\mathbf K}^{(\mu)}+\frac12 I\right)
 \end{aligned}
 \label{eq:entropy_single_mode}
\end{equation}
Here $2\varepsilon=\slash\!\!\! k_B$, so the complete retained-mode operator is
\begin{equation}
 \widehat\Sigma_M
 =\slash\!\!\! k_B\sum_{\mu=1}^{n}
 \sum_{\mathbf K\in\mathcal K_M}
 \left(N_{\mathbf K}^{(\mu)}+\frac12 I\right)
 \label{eq:entropy_number_operator}
\end{equation}
Write its number states as $|\{n\}\rangle$, where $\{n\}$ denotes the
complete collection $\{n_{\mathbf K}^{(\mu)}\}$ of $nM$ nonnegative
integer occupations. Then
\begin{equation}
 \begin{aligned}
 N_{\mathbf K}^{(\mu)}|\{n\}\rangle
 &=n_{\mathbf K}^{(\mu)}|\{n\}\rangle,\\
 \widehat\Sigma_M|\{n\}\rangle
 &=\Sigma_{\{n\}}|\{n\}\rangle
 \end{aligned}
 \label{eq:entropy_number_eigenstates}
\end{equation}
The total excitation number and the eigenvalue before matching are
\begin{equation}
 \begin{aligned}
 N&=\sum_{\mu=1}^{n}\sum_{\mathbf K\in\mathcal K_M}
                n_{\mathbf K}^{(\mu)},\\
 \Sigma_{\{n\}}&=\slash\!\!\! k_B N+\frac{nM\slash\!\!\! k_B}{2}
 \end{aligned}
\end{equation}
To recover Prigogine's equilibrium entropy production fluctuation theorem $nk_B/2$ \cite{Prigogine1968, epq-prigogine-bio}, we have to impose the finite-mode matching condition \cite{Sonnino2024a}:
\begin{equation}
 \frac{nM\slash\!\!\! k_B}{2}=\frac{nk_B}{2},
 \qquad M\slash\!\!\! k_B=k_B
 \label{eq:entropy_cutoff_matching}
\end{equation}
This is the correspondence condition in \cite{Sonnino2024a}: the entropy ground value is
$nM\slash\!\!\! k_B/2$ and the cutoff is $M=k_B/\slash\!\!\! k_B$.
Here $M=k_B/\slash\!\!\! k_B$ must be a positive integer for an exact sharp
cutoff of equally weighted independent modes. With the commutator scale $\varepsilon=\slash\!\!\! k_B/2$, the quadratic
prefactor in Eq.~\eqref{eq:entropy_quadratic_operator} is $1$.
A prefactor $1/2$ in these same variables would instead give a spacing
$\varepsilon$ and an incorrect factor of two in the matching condition.
The scale in the commutator and the spacing of the entropy levels must
therefore be distinguished. With this matching, the general spectrum is
\begin{equation}
\Sigma_N=\slash\!\!\! k_B N+\frac n2 k_B,
        \qquad N\in\mathbb N_0
 \label{eq:entropy_general_spectrumT}
\end{equation}
Its minimum and adjacent-level spacing are
\begin{equation}
 \Sigma_{\min}=\frac n2 k_B,
 \qquad \Delta\Sigma=\slash\!\!\! k_B=2\varepsilon
 \label{eq:entropy_spacing}
\end{equation}
For unrestricted occupations of the $nM$ independent equally weighted
oscillators, the level multiplicity is $\binom{N+nM-1}{nM-1}$. The ground state has zero occupations and
nonzero canonical second moments; its eigenvalue $nk_B/2$ is an
expectation of the quadratic observable, rather than its variance.
The finite cutoff makes the zero-point term finite, while the
occupation numbers and hence the excitation ladder remain unbounded.

\subsection{Two coupled thermodynamic modes}

The electrical and thermal transport channels have $n=2$. Each retained
$\mathbf K$ therefore contributes two canonical pairs, so the example
contains $2M$ independent oscillators. Specializing the general operator
gives
\begin{equation}
 \begin{aligned}
 \widehat\Sigma_M
 &=\sum_{\mathbf K\in\mathcal K_M}
 \sum_{\mu=1}^{2}
 \left[(q_{\mathbf K}^{(\mu)})^2
      +(p_{\mathbf K}^{(\mu)})^2\right]\\
 &=\slash\!\!\! k_B\sum_{\mathbf K\in\mathcal K_M}
 \sum_{\mu=1}^{2}
 \left(N_{\mathbf K}^{(\mu)}+\frac12 I\right)
 \end{aligned}
 \label{eq:entropy_two_mode_operator}
\end{equation}
For the two-force example, define the total excitation number by
\begin{equation}
 N=\sum_{\mathbf K\in\mathcal K_M}
 \sum_{\mu=1}^{2}n_{\mathbf K}^{(\mu)},
 \qquad N\in\mathbb N_0
\end{equation}
The entropy-production eigenvalues are consequently
\begin{equation}
 \begin{aligned}
 \Sigma_{\{n\}}
 &=\slash\!\!\! k_B\sum_{\mathbf K\in\mathcal K_M}
 \sum_{\mu=1}^{2}
 \left(n_{\mathbf K}^{(\mu)}+\frac12\right)\\
 &=\slash\!\!\! k_B(N+M)
 \end{aligned}
 \label{eq:entropy_two_mode_spectrum}
\end{equation}
Using Eq.~\eqref{eq:entropy_cutoff_matching} gives the levels
\begin{equation}
\Sigma_N=k_B+\slash\!\!\! k_B N,\qquad N=0,1,2,\ldots
 \label{eq:entropy_total_levels}
\end{equation}
Thus the ground level, adjacent-level spacing, and excess above the
reference are
\begin{equation}
 \begin{aligned}
 \Sigma_{\min}&=k_B,\\
 \Delta\Sigma&=\slash\!\!\! k_B=2\varepsilon,\\
 \Sigma_{\mathrm{ex},N}&:=\Sigma_N-k_B=\slash\!\!\! k_B N
 \end{aligned}
 \label{eq:entropy_two_force_spacing}
\end{equation}
The first few levels are explicitly
\begin{equation}
 \begin{aligned}
 \Sigma_0&=k_B,\\
 \Sigma_1&=k_B+\slash\!\!\! k_B,\\
 \Sigma_2&=k_B+2\slash\!\!\! k_B,\\
 \Sigma_3&=k_B+3\slash\!\!\! k_B
 \end{aligned}
 \label{eq:entropy_first_levels}
\end{equation}
Equivalently, their dimensionless values are
\begin{equation}
 \frac{\Sigma_N}{k_B}=1+\frac{N}{M}
\end{equation}
The degeneracy counts the ways to distribute $N$ identical excitation
quanta among $2M$ labeled independent oscillators:
\begin{equation}
d_N=\binom{N+2M-1}{2M-1}
 \label{eq:entropy_degeneracy}
\end{equation}
In particular, a single excitation can occupy either principal channel
at any of the $M$ retained labels, giving $d_1=2M$. At $N=2$, two quanta
can occupy one oscillator or one quantum can occupy each of two distinct
oscillators, giving $d_2=2M+\binom{2M}{2}=M(2M+1)$.
Table~\ref{tab:entropy_two_force_levels} displays the first levels.
\begin{table}[t]
\caption{Entropy-production levels for two thermodynamic forces and $M$
independent retained modes per channel, with $M\slash\!\!\! k_B=k_B$.
The normalized excess is $(\Sigma_N-k_B)/\slash\!\!\! k_B=N$.}
\label{tab:entropy_two_force_levels}
\centering
\begin{ruledtabular}
\begin{tabular}{ccc}
$N$ & $\Sigma_N/k_B$ & $d_N$\\
0 & $1$       & $1$\\
1 & $1+1/M$   & $2M$\\
2 & $1+2/M$   & $M(2M+1)$\\
3 & $1+3/M$   & $\binom{2M+2}{3}$\\
4 & $1+4/M$   & $\binom{2M+3}{4}$
\end{tabular}
\end{ruledtabular}
\end{table}
The simpler formula $d_N=N+1$ applies only when $M=1$, in which case
there are exactly two oscillators and the matching condition gives
$\slash\!\!\! k_B=k_B$. For general $M$, writing the state using only two
occupations $n_1,n_2$ hides the retained-mode multiplicity. One may
instead define channel totals
\begin{equation}
 N_\mu=\sum_{\mathbf K\in\mathcal K_M}n_{\mathbf K}^{(\mu)},
 \qquad N=N_1+N_2
\end{equation}
Then $\Sigma_{N_1,N_2}=k_B+\slash\!\!\! k_B(N_1+N_2)$, but each pair of
channel totals has multiplicity
\begin{equation}
 g_{N_1,N_2}
 =\binom{N_1+M-1}{M-1}\binom{N_2+M-1}{M-1}
\end{equation}
Summing this multiplicity over $N_1+N_2=N$ recovers
Eq.~\eqref{eq:entropy_degeneracy}. For the $\mathrm{Bi_2Te_3}$ film, the previously calculated coupling is
$r\simeq-0.636$. With the identification $\ell=\widetilde L$ in
normalized coordinates, its principal directions are
$e_\pm=(1,\pm1)^T/\sqrt2$ and its eigenvalues are
$\lambda_\pm=1\pm r$. At each retained label, the canonical reduction
has the form
\begin{equation}
 q_{\pm,\mathbf K}=\frac{\xi_{\pm,\mathbf K}}{\sqrt{1\pm r}},
 \qquad
 p_{\pm,\mathbf K}=\sqrt{1\pm r}\,Y_{\pm,\mathbf K}
\end{equation}
Here the principal components of $\bm\xi_{\mathbf K}$ and
$\bm Y_{\mathbf K}$ are obtained by projection onto $e_\pm$.
Numerically, the coordinate factors are approximately $1.658$ and
$0.782$, while their conjugate factors are $0.603$ and $1.279$.
The reciprocal coefficients cancel in the normalized quadratic form,
so the coupling changes the principal variables without splitting the
levels in Eq.~\eqref{eq:entropy_total_levels}. Their spacing is the
mesoscopic entropy quantum $\slash\!\!\! k_B$, rather than
an eigenvalue of the transport matrix or spectral metric. A numerical
value of $\slash\!\!\! k_B$, or equivalently $M$, makes the level spacing
numerical; the dimensionless transport coupling alone does not supply
that value.

\subsection{Distinction from the evolution spectrum}

The entropy spectrum above must also be distinguished from the
entropy-production rate. The rate operator carries mode frequencies \cite{Sonnino2024a}:
\begin{equation}
 \widehat\sigma_T
 =\sum_{\mu=1}^n\sum_{\mathbf K\in\mathcal K_M}
 \slash\!\!\! k_B\omega_{\mathbf K}
 \left(N_{\mathbf K}^{(\mu)}+\frac12 I\right)
\end{equation}
Its excitation increments are $\slash\!\!\! k_B\omega_{\mathbf K}$.
The equally spaced entropy ladder for $\Sigma$ does not by itself
imply equal spacing of this rate spectrum when the frequencies differ. The entropy-production observable and the evolution generator are
different operators. In the constant Onsager regime,
\begin{equation}
 D=-\frac12L_{\mu\nu}\partial^\mu\partial^\nu
 \label{eq:entropy_evolution_comparison}
\end{equation}
For positive definite $L$ on $\mathbb R^n$, Fourier transformation gives
\begin{equation}
 D(k)=\frac12L_{\mu\nu}k_\mu k_\nu
 \label{eq:entropy_D_fourier}
\end{equation}
The continuous Fourier parameter $k$ therefore yields
\begin{equation}
 \operatorname{spec}(D)=[0,\infty)
 \label{eq:entropy_D_spectrum}
\end{equation}
This statement concerns the unbounded-domain generator. It assumes
that the finite set $\mathcal K_M$ labels the retained canonical
oscillators without also restricting this generator's Fourier domain.
If the same cutoff is applied to $D$, or spatial boundary conditions
are imposed, its spectrum must be evaluated on that restricted domain;
the unrestricted result above must not be carried over automatically. For the truncated entropy-production observable with $n$ channels,
\begin{equation}
 \begin{aligned}
 \Sigma_{\{n\}}
 &=\slash\!\!\! k_B\sum_{\mu=1}^n
 \sum_{\mathbf K\in\mathcal K_M}n_{\mathbf K}^{(\mu)}
 +\frac n2 k_B,\\
 \Sigma_{\min}&=\frac n2k_B,
 \qquad M\slash\!\!\! k_B=k_B
 \end{aligned}
 \label{eq:entropy_Sigma_spectrum}
\end{equation}
Thus, the two-force case has the discrete spectrum
\begin{equation}
 \operatorname{spec}(\widehat\Sigma_M)
 =\{k_B+\slash\!\!\! k_B N:N\in\mathbb N_0\}
 \label{eq:entropy_minimum_kb}
\end{equation}
The finite cutoff makes the zero-point sum finite, while the occupation
numbers remain unbounded. Hence the number of retained oscillators is
finite, but the ladder of entropy-production levels is infinite. Neither the factorization nor the Clifford--Dirac identity,
\begin{equation}
 D=T^\dagger T,\qquad \mathcal D^2=DI
 \label{eq:entropy_factorization_reminder}
\end{equation}
identifies $D$ with $\widehat\Sigma_M$. The confining quadratic
observable produces the oscillator levels, and the matching condition
fixes their equilibrium offset to $nk_B/2$. These levels remain
distinct from the evolution spectrum and from the spectral distances
calculated for the thermoelectric sample.

% ============================================================
% CONCLUSION
% ============================================================

\section{Conclusions}\label{sec:conclusions}

Starting from canonical relations between thermodynamic coordinates and their conjugate forces, we have constructed a coherent evolution generator compatible with mesoscopic flux observables and derived a distance from that same generator. The connection is explicit: transport fixes the differential coefficients of the evolution equation, while an exact first-order factorization determines the metric through Connes' spectral-distance prescription. The calculation accommodates coordinate-dependent coefficients under the stated positivity and regularity assumptions. This result extends the earlier canonical mesoscopic formulation, which introduced the commutation relations and investigated entropy discretization and the reduced constant $\slash\!\!\! k_B$ \cite{Sonnino2024a,Sonnino2024b,Sonnino2025}. Related equilibrium approaches also promote thermodynamic variables to operators and derive Schr\"odinger-like equations \cite{Santos2026}. Here we connect evolution in physical time to specified transport fluxes and use the resulting operator to reconstruct thermodynamic distance. The advance lies in establishing this joint construction, including the coefficient-derivative terms needed when transport varies across thermodynamic-coordinate space. Requiring each coordinate operator to evolve according to its corresponding self-adjoint flux determines the generator up to an additive real multiplication operator $U(\alpha)$. Choosing $U=0$ gives the minimal realization studied here. Its Schr\"odinger-type equation governs mesoscopic thermodynamic amplitudes, with $\slash\!\!\! k_B$ setting the evolution scale. The generator is self-adjoint, and the evolution is unitary. Irreversible relaxation therefore remains a separate physical problem, requiring additional mechanisms such as environmental coupling or coarse-graining. Factoring this generator yields a metric equal to twice the inverse of the symmetric transport matrix. Derivatives of the antisymmetric tensor can change the evolution without changing the distance. Transport thus carries dynamical information that the metric alone does not retain. This separation holds for variable coefficients, not just in the constant Onsager limit. In that limit, a Clifford--Dirac representation recovers the same metric and expresses its algebra through spinorial components without introducing physical spin. The bismuth telluride example illustrates what the geometric result means for transport data. At $ZT\simeq0.68$, equal displacements along the principal directions of normalized charge--heat coordinates have distances differing by a factor of approximately $2.12$. These distances describe thermodynamic transfers rather than physical lengths. Their reconstruction gives a geometric interpretation of the measured transport coefficients; it does not independently predict material performance or establish an observable effect of $\slash\!\!\! k_B$. Effects associated specifically with the reduced constant require a different test. The canonical relations impose uncertainty bounds, while the separate quadratic entropy-production observable has adjacent entropy levels separated by $\slash\!\!\! k_B$ under the stated retained-mode and equilibrium matching assumptions. We have made the mode counting explicit and distinguished this discrete spectrum from the evolution spectrum and from the entropy-production rate. The spacing is a conditional prediction of that observable, rather than a consequence of the geometric factorization. Assessing its observability requires a measurement prescription, an appropriate estimate of the reduced scale, and sufficient experimental resolution. Several questions remain open. An open-system extension could investigate how the coherent mesoscopic dynamics leads to effective relaxation. The reconstructed metric defines its own Levi-Civita connection, but additional physical input is needed to determine the generally non-Levi-Civita affine connection of Thermodynamic Field Theory. Neither issue alters the transport-derived metric established here; both matter for extending the framework to a fuller description of irreversible processes. The work therefore supplies a transport-compatible canonical evolution law and an exact route from its generator to thermodynamic distance. It also specifies where a possible signature of $\slash\!\!\! k_B$ enters and which additional assumptions that prediction requires. This formulation allows coherent evolution, geometric structure, and entropy-level effects to be investigated within one model without treating them as interchangeable consequences of the canonical postulate.

% ============================================================
% REFERENCES
% ============================================================

\end{document}